\documentclass[a4paper]{spie}  
\usepackage{cite}
\usepackage{amsmath,amssymb,amsfonts}
\usepackage{textcomp}
\usepackage{xcolor}

\usepackage{graphicx}
\usepackage{bm}
\usepackage{comment}
\usepackage[subrefformat=parens]{subcaption}

\title{Privacy-Preserving Machine Learning Using EtC Images} 

\author{Ayana Kawamura, Yuma Kinoshita, Hitoshi Kiya
\skiplinehalf
Tokyo Metropolitan University, Tokyo, Japan
}

\begin{document} 
\newcommand{\red}[1]{\textcolor{black}{#1}}
  \maketitle 

\begin{abstract}
In this paper, we propose a novel privacy-preserving machine learning scheme with encrypted images, called EtC (Encryption-then-Compression) images.
Using machine learning algorithms in cloud environments has been spreading in many fields.
However, there are serious issues with it for end users, due to semi-trusted cloud providers.
Accordingly, we propose using EtC images, which have been proposed for EtC systems with JPEG compression.
In this paper, a novel property of EtC images is considered under the use of z-score normalization.
It is demonstrated that the use of EtC images allows us not only to protect visual information of images, but also to preserve both the Euclidean distance and the inner product between vectors.
In addition, dimensionality reduction is shown to can be applied to EtC images for fast and accurate matching.
In an experiment, the proposed scheme is applied to a fac\red{ial} recognition algorithm with classifiers for confirming the effectiveness of the scheme under the use of support vector machine (SVM) with the kernel trick. 
\end{abstract}



\section{INTRODUCTION}
\label{sec:intro}  

Cloud computing and edge computing have been spreading in many fields with the development of cloud services.
However, the computing environment has \red{some} serious issues for end users, such as \red{the} unauthorized use \red{of services, data leaks}, and privacy \red{being} compromise\red{d} due to unreliab\red{le} providers and \red{some} accidents\cite{huang2014survey}.
 Various methods have been proposed for privacy-preserving computation.
The methods are classified into two types: perceptual encryption-based type\cite{Chuman2018,Warit2019APSIPA,Itier2019,Tanaka2018} and homomorphic encryption (HE)-based one\cite{Phong2018,Wang2018,Yang2017,Saxe2018}.

 In recent years, considerable efforts have been made in the fields of fully HE and multi-party computation\cite{Araki2017}.
Some attempts with HE-based one have been made \red{on} learning\cite{Phong2018,Wang2018,Yang2017,Saxe2018}.
However, HE-based schemes require algorithms specialized for computing encrypted data \red{be prepared}, and high computational complexity\cite{Wang2018}.
In addition, it is also difficult to maintain high computational accuracy\cite{Yang2017}.

 Because of \red{this}, we propose a privacy-preserving machine learning scheme based on a perceptual encryption method in this paper.
We focus on encrypted images, called \red{"e}ncryption-then-\red{c}ompression (EtC) images,\red{"} which have been proposed for EtC systems with JPEG compression\cite{Chuman2018,Warit2019APSIPA,kurihara2015encryption,KURIHARA2015,Kuri_2017,watanabe2015encryption}.
So far, the safety of the EtC systems has been evaluated \red{on the basis of} the key space \red{under the assumption of} brute-force attacks, and robustness against jigsaw puzzle attacks has been discussed \cite{CHUMAN2017ICASSP,Warit2018}.
In this paper, a novel property of EtC images is considered under the use of z-score normalization.
The novel property allows us to securely compute machine learning algorithms without any degradation \red{in} performances.
In addition, dimensionality reduction is shown to \red{make it possible} to apply EtC images for fast and accurate matching of visual features.
In an experiment, the proposed scheme is applied to a fac\red{ial} recognition algorithm with classifiers to confirm the effectiveness of the scheme under the use of support vector machine (SVM).

\section{EtC image} 

\begin{figure}[t]
\centering
  \begin{tabular}{c c}
  \begin{minipage}[b]{0.5\hsize}
      \begin{center}
 	 \centering\includegraphics[width=9cm]{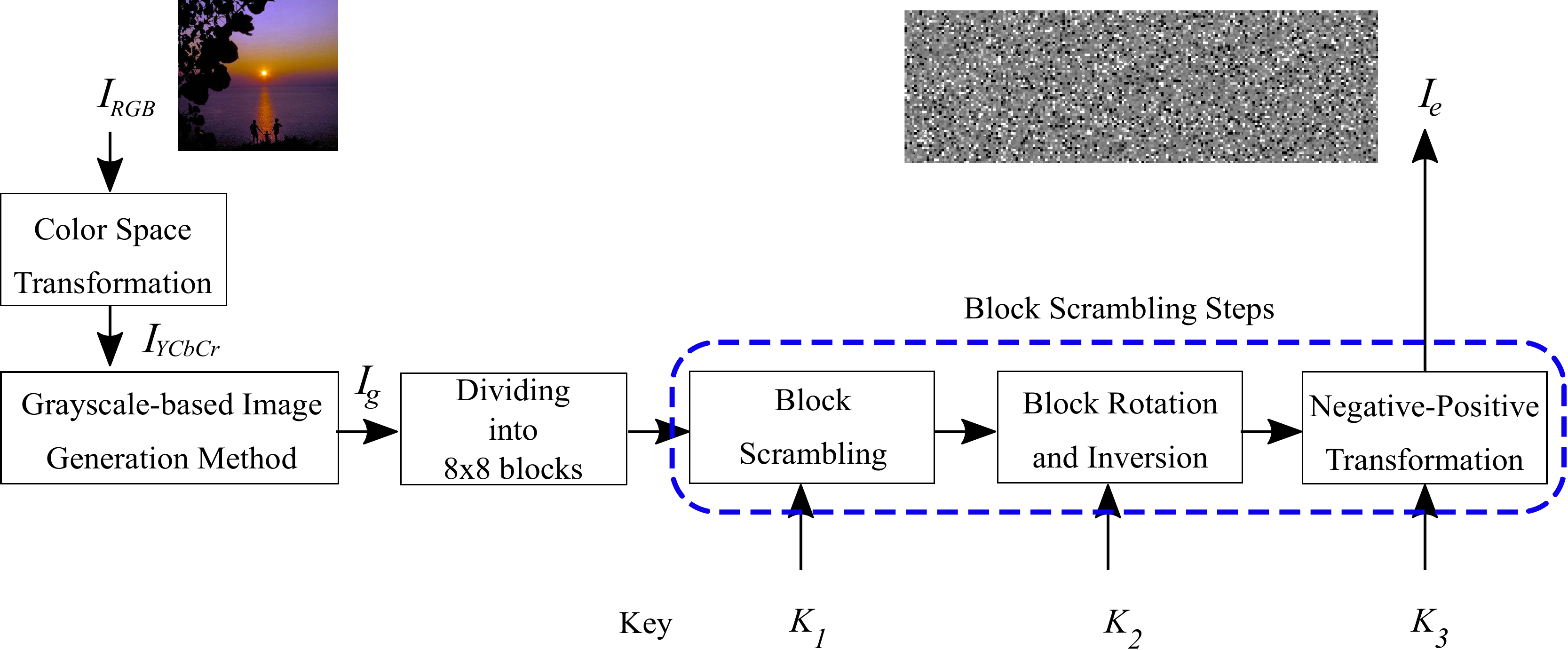}
      \end{center}
\caption{Encryption steps for generating EtC images\cite{Chuman2018,Warit2019APSIPA}}
\label{enc}
    \end{minipage}
    &
     \begin{minipage}[b]{0.5\hsize}
      \begin{center}
        \centering\includegraphics[width=7.5cm]{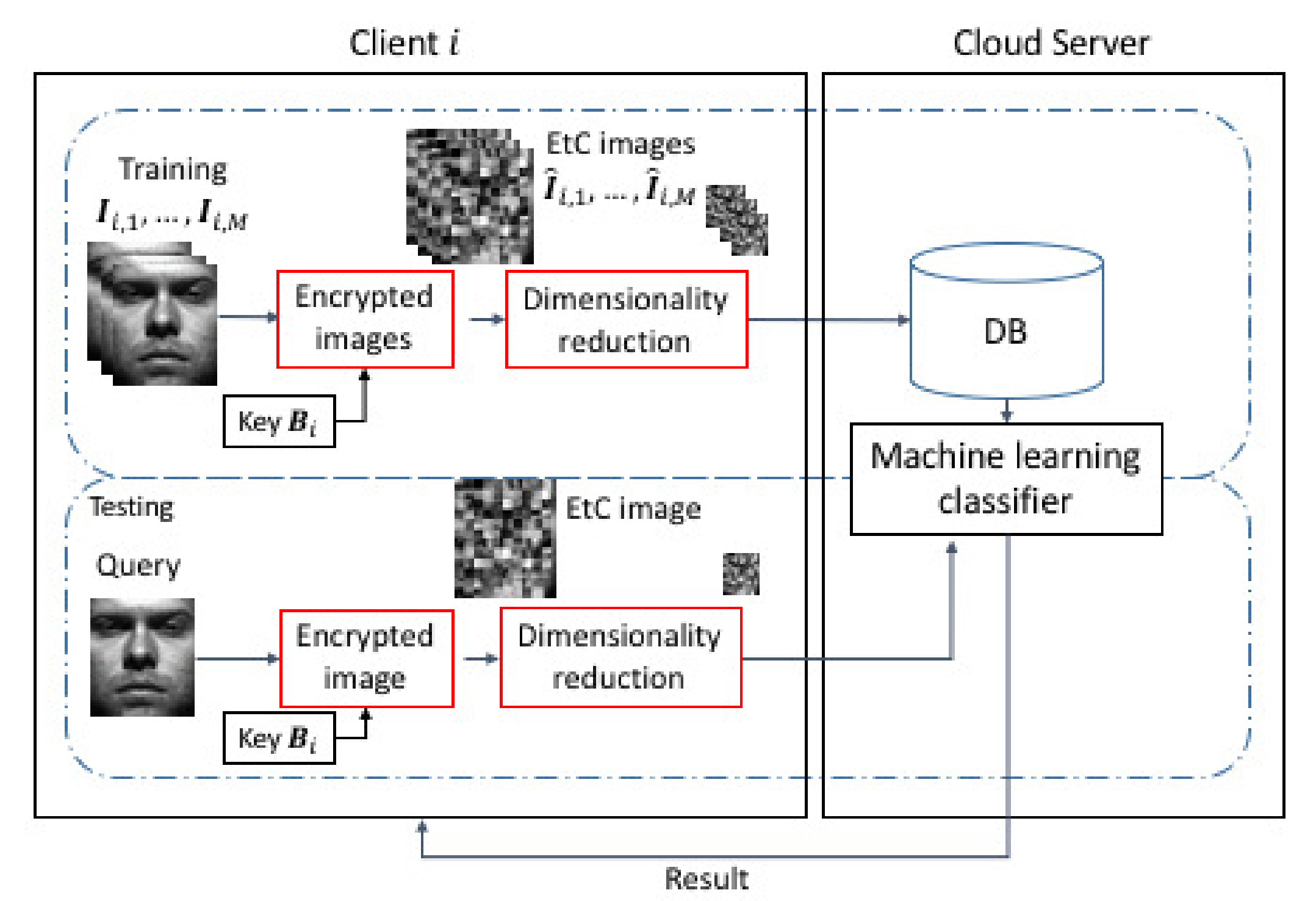}
      \end{center}
\caption{Privacy-preserving machine learning}
\label{svmsystem}
    \end{minipage}
\end{tabular}
\end{figure}

 As mentioned above, we focus on EtC images.
EtC ones are \red{robust} again\red{st} various ciphertext-only attacks\red{,} including jigsaw puzzle solver attacks\cite{CHUMAN2017ICASSP,Warit2018}.
In this paper, we show that EtC images have a novel property, and \red{this} property allows us to carry out privacy-preserving machine learning.

 In Figure \ref{enc}\cite{Chuman2018,Warit2019APSIPA}, the procedure of generating EtC images is shown.
In this paper, images encrypted \red{with} the above steps are referred to as EtC images.

\section{PROPOSED PRIVACY-PRESERVING MACHINE LEARNING} \label{sec:sections}

\subsection{Scenario}
 The procedure of the proposed scheme is summarized here.

\noindent A. Model Training (See Fig.\ref{svmsystem}): 1) Prepare EtC images by using a block-based encryption method \cite{Chuman2018} in a client, and send them to a cloud server.
2) Perform dimensionality reduction in the encrypted domain, followed by training a model with the reduced data in the encrypted domain.

\noindent B. Testing: 1) Send an EtC image as a test one to the cloud server.
2) Solve a classification problem with the trained model, after the dimensionality reduction of the image as well as for training ones.
3) Return a result to the client.

Note that the cloud provider has no both visual information of images and the secret key. In addition, the provider can carry out directly machine learning algorithms with encrypted data.
The performance of the proposed scheme will be demonstrated to be the same as that of plain images under the use of z-score normalization.

\subsection{Novel property of EtC images}

In the encryption steps in Fig.\ref{enc}, block scramble, block rotation and
inversion are carried out for permuting pixels.
These operations can be represented as an orthogonal matrix.
As a result, block scramble, block rotation and inversion preserve both the Euclidian distance and the inner product between vectors.
In addition, when applying z-score normalization to EtC images, its normalization allows us to maintain the inner product between vectors even when an negative-positive transformation is carried out.
Therefore, EtC images can provide exact the same as the performance of plain images under the use of typical machine learning algorithms with the kernel trick, including k-nearest neighbor, random forest, the radial basis function (RBF) kernel and the polynomial kernel.

\section{SIMULATION} 

 The proposed scheme was applied to fac\red{ial} recognition experiments \red{that} were carried out with a SVM algorithm.

\subsection{Simulation condition}
\begin{figure}[t]
\centering
  \begin{tabular}{c c c c}
     \begin{minipage}[b]{0.15\hsize}
 	     \centering\includegraphics[width = 1.5cm]{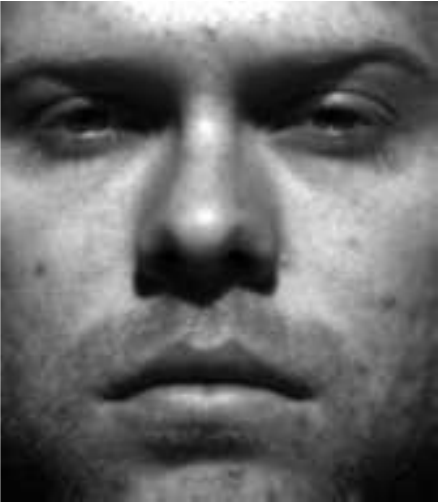}
     \end{minipage}
      &
     \begin{minipage}[b]{0.15\hsize}
        \centering\includegraphics[width = 1.5cm]{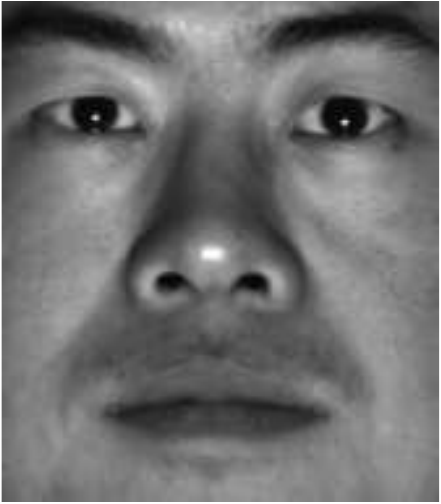}
    \end{minipage}
      &
    \begin{minipage}[b]{0.15\hsize}
   	 \centering\includegraphics[width = 1.4cm]{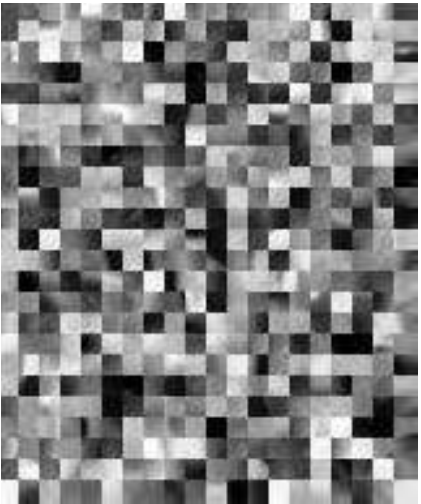}
    \end{minipage}
      &
    \begin{minipage}[b]{0.15\hsize}
      \centering\includegraphics[width = 1.4cm]{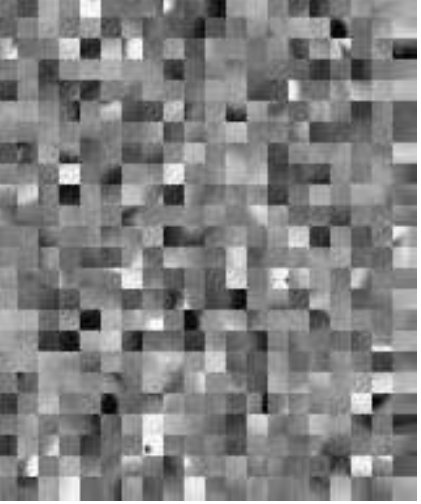}
    \end{minipage}
  \\
    \multicolumn{2}{c}{(a) Original image}
     &
    \multicolumn{2}{c}{(b) EtC image}
  \end{tabular}
\caption{\red{E}xample of facial images}
\label{encrypttemplate}
\end{figure}

 We used \red{the} Extended Yale Face Database B\cite{Geo2001}\red{, which} consists of 2432 frontal facial images with 192 $\times$ 160 pixels \red{for} 38 persons.
64 images for each person were divided into half randomly for training data samples and queries.
$8\times 8$ was used as a block size of encryption. 
Dimensionality \red{was reduced} with reduction ratio\red{s} of 1/20, 1/40, 1/60\red{, and} 1/80, so $192\times 160=30720$ dimensions \red{was} reduced to 1536 dimensions for \red{a ratio of} 1/20.
After the dimensionality reduction, we applied z-score normalization to the reduced vectors.
Figure \ref{encrypttemplate} shows an example of original images and the EtC images.
SVM was used as an example of machine learning algorithms, with \red{the} RBF kernel and linear kernel.

\vspace{-0.2in}
\subsection{Simulation results}
In this experiment, \red{the f}alse \red{r}eject \red{r}ate (FRR), \red{f}alse \red{a}ccept \red{r}ate (FAR), and \red{e}qual \red{e}rror \red{r}ate (EER)\red{,} at which FAR is equal to FRR\red{,} were used to evaluate the performance.

\begin{figure}[t]
\centering
  \begin{tabular}{cc}
  \begin{minipage}[b]{0.5\hsize}
    \centering
    \includegraphics[width = 5cm]{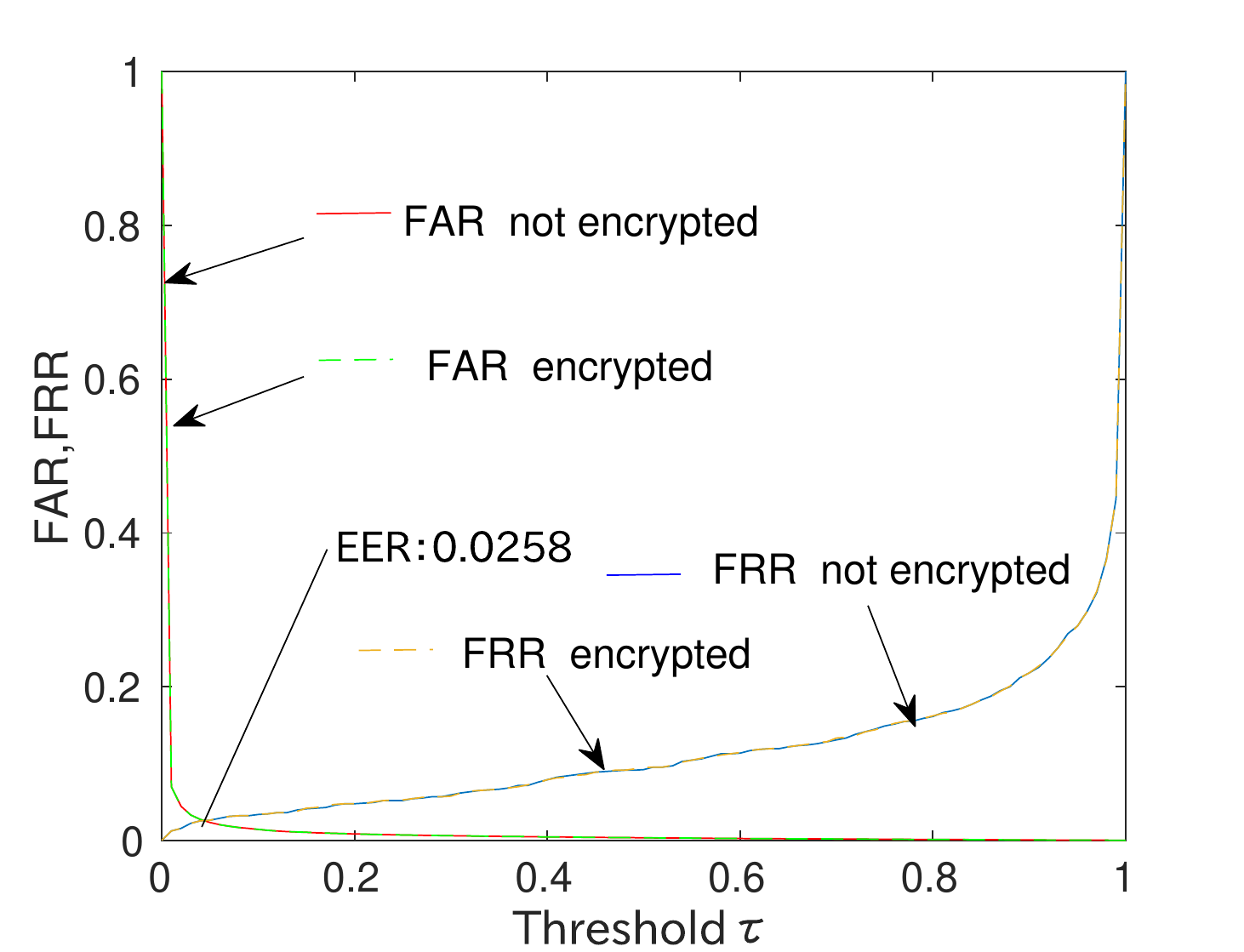}\\
    (a) Linear kernel
  \end{minipage}
    &
    \begin{minipage}[b]{0.5\hsize}
        \centering
        \includegraphics[width = 5cm]{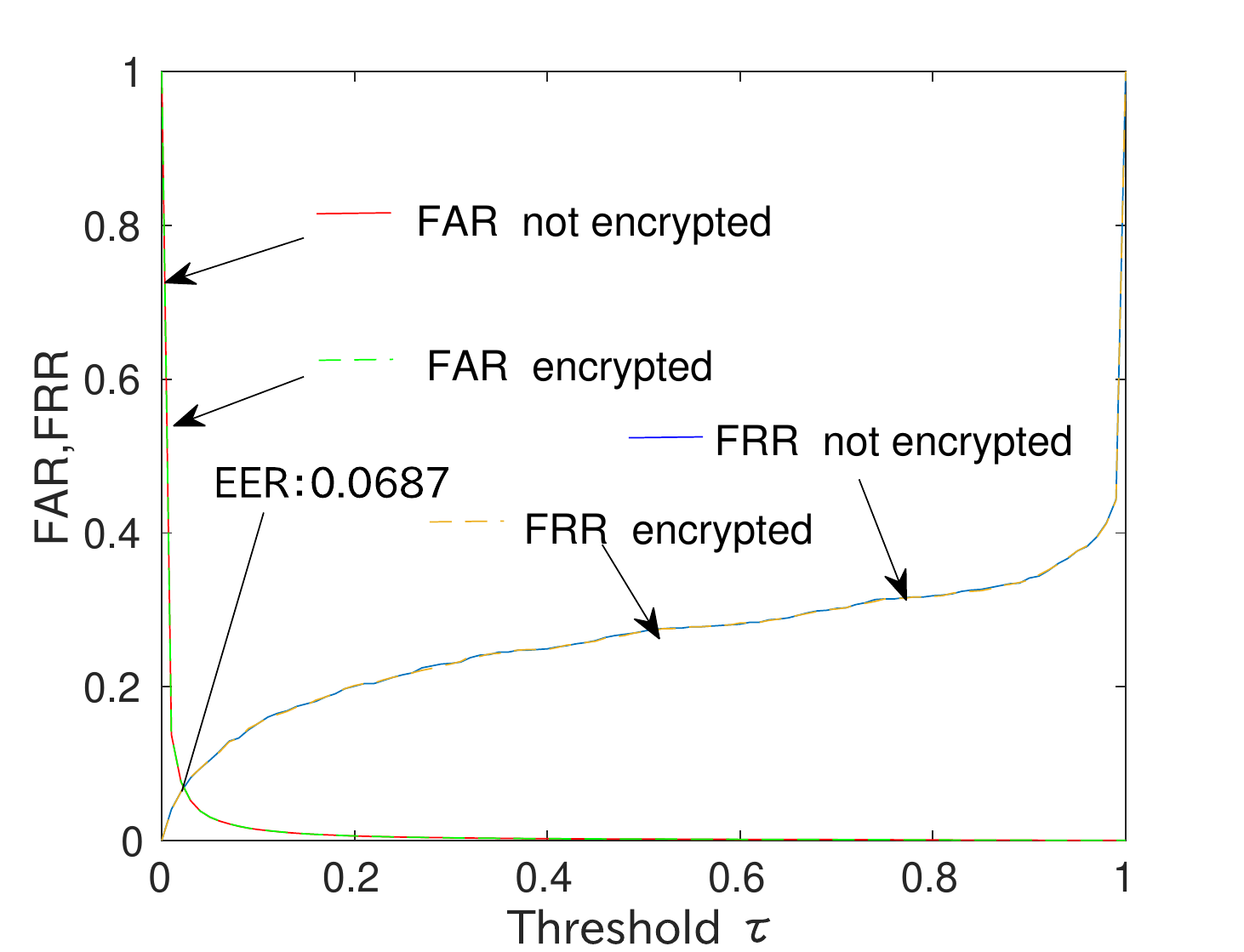}\\
      (b) RBF kernel
    \end{minipage}
\end{tabular}
\centering\caption{FRR and FAR for SVM \protect\linebreak ($\bm{B}_1=\bm{B}_2=\ldots=\bm{B}_N$, reduction ratio = 1/64)}
\label{result}
\end{figure}

\begin{table}[t]
\begin{tabular}{cc}
\begin{minipage}[c]{0.5\hsize}
\centering
\caption{EER values for SVM with linear kernel \protect\linebreak ($\bm{B}_1\neq \bm{B}_2\neq \ldots \neq \bm{B}_N$)}
\label{linearkeydif}
\begin{tabular}{|c|c|c|}
\hline
\begin{tabular}{c}reduction\\ ratio\end{tabular} & \begin{tabular}{c}not\\ encrypted\end{tabular} & encrypted \\ \hline
1/20 & 0.0223 & 0.000744 \\ \hline
1/40 & 0.0247 & 0.000835 \\ \hline
1/60 & 0.0271 & 0.000777 \\ \hline
1/80 & 0.0296 & 0.000779 \\ \hline
\end{tabular}
\end{minipage} &
\begin{minipage}[c]{0.5\hsize}
\centering
\caption{EER values for SVM with RBF kernel \protect\linebreak ($\bm{B}_1\neq \bm{B}_2\neq \ldots \neq \bm{B}_N$)}
\label{RBFkeydif}
\begin{tabular}{|c|c|c|}
\hline
\begin{tabular}{c}reduction\\ ratio\end{tabular} & \begin{tabular}{c}not\\ encrypted\end{tabular} & encrypted \\ \hline
1/20 & 0.0504 & 0.000448 \\ \hline
1/40 & 0.0644 & 0.00112 \\ \hline
1/60 & 0.0732 & 0.00779 \\ \hline
1/80 & 0.0863 & 0.00855 \\ \hline
\end{tabular}
\end{minipage}
\end{tabular}
\end{table}

\subsubsection{Key condition 1: $\bm{B}_1=\bm{B}_2=\ldots=\bm{B}_N$}
 Experimental results are shown in Figure \ref{result} in the case of using a common key for all clients (Key condition 1).
The results demonstrate that SVM classifiers with encrypted images (encrypted in Fig.\ref{result}) \red{performed} the same as SVM classifiers with the original images (not encrypted in Fig.\ref{result}).
From the results, the proposed \red{scheme was} confirmed to give no influence \red{on} the performance of SVM classifiers under key condition 1 and z-score normalization.

\subsubsection{Key condition 2: $\bm{B}_1\neq \bm{B}_2\neq\ldots\neq \bm{B}_N$}
 Table\red{s} \ref{linearkeydif} and \ref{RBFkeydif} show results in the case of using a different key for each client (key condition 2).
In this condition, it is expected that a query is authenticated only when it meets two requirements, i.e.\red{,} the same key and the same person as for training the model, although only the same person \red{is} required under key condition 1.
Therefore, the performance \red{with} encrypted \red{images is slightly} different from \red{that} of plain \red{images}, so the EER values of \red{"}encrypted\red{"} in Tables \ref{linearkeydif} and \ref{RBFkeydif} were smaller than those of \red{"}not encrypted\red{"} due to the strict requirements.
As a result, the EER values under key condition 2 outperformed those under key condition 1.

\section{CONCLUSION} \label{sec:misc}

 We proposed a privacy-preserving machine learning scheme with EtC images.
A novel property allows us to apply EtC images to some machine learning algorithms without any degradation \red{in} classification performance.
Moreover, two key conditions were considered to enhance robustness against various attacks.
A number of fac\red{ial} recognition experiments using SVM were also demonstrated to show the effectiveness of the proposed \red{scheme}.

 



\end{document}